# Growth, Crystal Structure and Magnetic Characterization of Zn-Stabilized CePtIn$_4$

Elizabeth M. Carnicom[1,^], Tomasz Klimczuk[2], Fabian von Rohr[1], Michal J. Winiarski[2], Tai Kong[1], Karoline Stolze[1], Weiwei Xie[3], Satya K. Kushwaha[1], and Robert J. Cava[1,^]

[1]Department of Chemistry, Princeton University, Princeton, New Jersey 08544
[2]Department of Physics, Gdansk University of Technology, Gdansk Poland 80-233
[3]Department of Chemistry, Louisiana State University, Baton Rouge LA 70803

(Date Posted: May 12, 2017)

The growth and characterization of CePtIn$_4$, stabilized by 10% Zn substitution for In, is reported. The new material is orthorhombic, space group *Cmcm* (No. 63), with lattice parameters $a$ = 4.51751(4) Å, $b$ = 16.7570(2) Å, and $c$ = 7.36682(8) Å, and the refined crystal composition has 10% of Zn substituted for In, i.e. the crystals are CePt(In$_{3.6}$Zn$_{0.1}$)$_4$. Crystals were grown using a self-flux method: only growths containing Zn yielded CePtIn$_4$ crystals, while Ce$_3$Pt$_4$In$_{13}$ crystals formed when Zn was not present. Anisotropic temperature-dependent magnetic susceptibilities for single crystals show that Zn-stabilized CePtIn$_4$ orders magnetically at ~1.9 K. High-temperature Curie-Weiss fits indicate an effective moment of ~2.30 µ$_B$/ Ce and a directionally averaged Weiss-temperature of approximately - 9 K. Specific heat data shows a peak consistent with the ordering temperature seen in the magnetic susceptibility data. Zn-stabilized CePtIn$_4$ is metallic and displays no superconducting transition down to 0.14 K.

## 1. Introduction

Indium-rich ternary intermetallic compounds containing a rare-earth (*RE*) element and a transition metal (*T*) have been of interest for many years due to both their diverse electrical and magnetic properties.[1–5] More specifically, indides with general formula *RET*In$_4$ adopting the YNiAl$_4$-type structure have been more frequently studied[3,5–12] than the analogous gallides[13,14] and aluminides[15–17] with the same structure. EuNiIn$_4$, for example, is an antiferromagnet (AFM) with a Néel temperature T$_N$ = 16 K,[18] EuPtIn$_4$ displays complex magnetic behavior with AFM transitions near 13 K and 5.5 K,[4] and solid solutions of CeNiIn$_4$/CeNiAl$_4$ have potential application as thermoelectric materials.[19] When the *RE* element is cerium, heavy fermion behavior,[20] superconductivity,[21,22] or mixed valence behavior[23–25] is sometimes observed. The variety of ground states for cerium-based intermetallic compounds originates from the competition between RKKY interactions and the Kondo effect.[25] CeNiIn$_4$, the *T*=3*d* element analog of the compound studied in this work, orders antiferromagnetically below 1.4 K.[26] Only the crystal structure for the isostructural *T*=4*d* Pd analog CePdIn$_4$ has been reported,[28,29] and until this report there appear to be no publications describing the crystal structure and elementary properties of the *T*=5*d* Pt analog, CePtIn$_4$.

Here we report the crystal growth, crystal structure, and elementary magnetic and electrical transport properties of previously unreported CePtIn$_4$ (denominated as "CePtIn$_4$-Zn" in the following) stabilized as single crystals by 10% Zn substitution for In. Crystal growths under our conditions without the addition of Zn did not yield the desired CePtIn$_4$ phase but instead resulted in the growth of cubic Ce$_3$Pt$_4$In$_{13}$ crystals. Attempts to grow polycrystalline Zn-free CePtIn$_4$ by arc-melting were not successful, but the compound was seen as a major phase in high-pressure high-temperature synthesis. Crystals of CePtIn$_4$-Zn display a magnetic transition at low temperature. Resistivity, specific heat and direction-dependent magnetic susceptibility measurements on single crystals are reported.

## 2. Experimental

The starting materials for the synthesis of CePtIn$_4$-Zn were cerium (>99.9%, chunk, Aldrich, stored under oil), platinum powder (99.9%, 200 mesh, Alfa), indium tear drops (99.9%, Alfa), and zinc (shot, <12 mm, >99.9%, Aldrich). Fresh pieces of cerium were cut and stored in an argon-filled glove box. Crystals were grown using a flux method by using a 2:2:2:94 ratio of cerium, platinum, and zinc to indium. The starting materials were then added to a carbon-coated quartz tube with pieces of quartz wool and sealed under vacuum (~80mTorr). The samples were heated to 1100°C at a rate of 60°C/h, held at 1100°C for 0.3h, cooled to 750°C over 4h, then cooled to 450°C over 150h, and finally held at 450°C for 48h. Following the crystal growth process, the samples were subjected to centrifugation to separate the crystals from the flux at 450°C. The resulting crystals were brick-shaped (1.5mm x 2mm x 8mm) and shiny in appearance. Residual flux was removed mechanically and by etching the crystals in 7% HCl for 0.5h. The crystals are stable in air and do not decompose over time. The same compound was seen as a major but not pure phase in high pressure synthesis experiments in the Ce-Pt-In system performed in BN crucibles at 6GPa and 1000°C in a Rockland Research cubic anvil cell. These impure high pressure materials were not characterized.

The purity of the CePtIn$_4$-Zn crystal products obtained from the flux growths was checked on crushed crystals by room temperature powder X-ray diffraction (pXRD) using a Bruker D8 Advance Eco, Cu K$_\alpha$ radiation ($\lambda$ =1.5406 Å), equipped with a LynxEye-XE detector. The surface of the crystals and their chemical composition were examined using an FEI Quanta 250FEG scanning electron microscope (SEM) equipped with an Apollo-X SDD energy-dispersive spectrometer (EDS). EDS data were processed by means of standardless quantitative analysis using the EDAX TEAM$^{TM}$ software. A Rietveld refinement was performed using the FullProf Suite with Thompson-Cox-Hastings pseudo-Voigt peak shapes. The EDS measurements specifying the relative amounts of indium and zinc in the crystals were used as a starting point for the structure refinement, and the refined In to Zn ratio was fully consistent with the formula determined by EDS. Crystal structure drawings were produced using the program VESTA.[30]

Physical property measurements were performed on the flux grown single crystals using a Quantum Design Physical Property Measurement (PPMS) Dynacool equipped with VSM and resistivity options. The crystal directions were aligned using the face-index method with a Bruker Apex II single crystal diffractometer (Mo radiation, $\lambda$ = 0.7107 Å). Anisotropic magnetic susceptibility was measured by fixing single crystals of CePtIn$_4$-Zn onto a silica sample holder using GE-varnish with the *a*, *b*, and *c*, axis parallel to the applied magnetic field in zero-field cooled (ZFC) temperature-dependent magnetic susceptibility measurements from 1.68 K to 300 K. Inverse susceptibility plots were fitted to the Curie-Weiss law,

$$\chi - \chi_0 = \frac{C}{T - \theta_{CW}}, \quad (1)$$

where $\chi$ is the magnetic susceptibility, C is the Curie constant, T is temperature (K), $\theta_{CW}$ is the Curie-Weiss theta, and $\chi_0$ is the temperature-independent contribution to the susceptibility. The effective magnetic moment ($\mu_B$) per cerium atom was determined using

$$\frac{\mu_B}{Ce} = \frac{\sqrt{8C}}{1\ mol\ Ce/f.u.}. \quad (2)$$

Field-dependent magnetization measurements were collected on single crystals using the PPMS at 1.7 K with a field sweep from 0 to 9 T with each crystallographic axis aligned parallel to the applied field. Because CePtIn$_4$-Zn showed linear magnetization (M) with an applied magnetic field (H) to $\mu_0$H significantly larger than 1 T when measured at 10 K, the magnetic susceptibility ($\chi$) for the single crystal measurements was defined as $\chi$ = M/H at $\mu_0$H = 1 T. The PPMS was used to measure the temperature-dependent electrical resistivity ($\rho$) using a four-probe method using H20E epoxy silver to make the contacts. Measurements were taken from 300 to 1.8 K and then from 2.4 to 0.14 K using an ADR (adiabatic demagnetization refrigerator) attachment in the PPMS with current set to 1000 µA. Heat capacity measurements were performed at 0, 0.25, 3, 6 and 9 T using a Quantum Design PPMS system on a small (16 mg) crystal using a standard relaxation method.

## 3. Results and Discussion

Crystals taken from a flux growth were crushed and analyzed by powder X-ray diffraction (pXRD), which showed that the previously unreported phase CePtIn$_4$-Zn is orthorhombic (*Cmcm*, No. 63) and adopts the YNiAl$_4$-type structure. **Figure 1** shows a comparison between the observed diffraction pattern and the intensities generated by the structural model in the Rietveld fit. Crystals with a fresh surface exposed, analyzed using energy dispersive X-ray spectroscopy (EDS), showed that the elemental composition of the new Zn-stabilized CePtIn$_4$ phase is CePt(In$_{0.9}$Zn$_{0.1}$)$_4$. These data were used as a starting point for the Rietveld refinement, where four different models were tested for the distribution of the Zn: (1) randomly distributing the Zn over the three independent In sites in the structure, In1, In2, and In3, (2) putting all the Zn on the In1 site, (3) putting all Zn on the In2 site, and (4) putting all the Zn on the In3 site. The model that gave the best fit to the diffraction data was when all the Zn was on the In1 site.

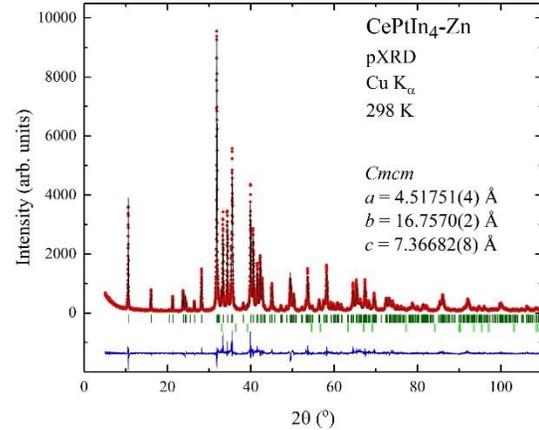

**Figure 1.** Room-temperature pXRD pattern of CePtIn$_4$-Zn. Bragg reflections are shown in green for CePtIn$_4$-Zn and light green for the impurity In. Observed data are shown in red, calculated data for the refined structure and formula of CePt(In$_{0.9}$Zn$_{0.1}$)$_4$ are shown in black, and the difference between the calculated and observed intensities is shown in blue.

The occupancy of mixed In1/Zn1 site was initially fixed to the appropriate values determined from elemental analysis, and all other structural parameters were allowed to freely refine. Once the refining values were stable, the occupancy of the In1/Zn1 site was then allowed to refine as well, so all parameters were freely varied to give the final values in **Table I**. The results of the refinement gave the formula CePt(In$_{0.9}$Zn$_{0.1}$)$_4$ for the crystals, fully consistent with the EDS data results. A small amount of indium impurity (<1.5 %) was present in the crushed crystals as a remnant from the flux growth and was included in the refinement.

The crystal structure of CePtIn$_4$-Zn viewed along the *a*-axis is shown in **Figure 2a**. The In, Zn, and Pt atoms form a polyanionic network surrounding the Ce atoms located in distorted hexagonal tunnels, and thus the new material can formally be described as Ce$^{3+}$[PtIn$_{3.6}$Zn$_{0.4}$]$^{3-}$.

**Table I. Crystal structure of Zn-stabilized CePtIn$_4$.** Space group *Cmcm* (No. 63), $a$ = 4.51751(4) Å, $b$ = 16.7570(2) Å, $c$ = 7.36682(8) Å, $V$ = 557.67(1) Å$^3$, $Z$ = 4.[a] The refined formula is CePt(In$_{0.9}$Zn$_{0.1}$)$_4$.

| Atom | Wyckoff Position | $x$ | $y$ | $z$ | $B_{iso}$ | Occupancy |
|---|---|---|---|---|---|---|
| Ce1 | 4$c$ | 0 | 0.3773(1) | ¼ | 3.01(5) | 1 |
| Pt1 | 4$c$ | 0 | 0.72551(8) | ¼ | 2.67(3) | 1 |
| In1 | 8$f$ | 0 | 0.18302(0) | 0.0501(2) | 2.41(2)[b] | 0.858(4) |
| Zn1 | 8$f$ | 0 | 0.18302(0) | 0.0501(2) | 2.41(2)[b] | 0.142(4) |
| In2 | 4$c$ | 0 | 0.5659(1) | ¼ | 2.41(2)[b] | 1 |
| In3 | 4$a$ | 0 | 0 | 0 | 2.41(2)[b] | 1 |

[a] $\chi^2$ =3.33; $wR_p$ = 11.2%; $R_p$ = 8.60%; $R(F^2)$ = 6.32%. [b] Values fixed such that In1, Zn1, In2, and In3 had the same thermal parameters (see text). Fractional make-up of the crushed single crystal sample used for the diffraction experiment: CePt(In$_{0.9}$Zn$_{0.1}$)$_4$, 98.7(4) %; In, 1.3(4) %.

(Our magnetic susceptibility study, described below, shows that the Ce is in the 3+ formal oxidation state). This is a common structural motif of indium-rich rare earth intermetallics with the YNiAl$_4$-type structure.[3,6,10,12,31] **Figure 2b** shows a picture of a typical brick-shaped crystal taken from the flux growth. According to the single crystal measurements indexing each face (**Figure 2c**), the longest physical dimension of the crystals is along the *a*-axis, and the shortest physical dimension of the crystals is along the *b*-axis. For flux-based crystal growths with Zn excluded, cubic Ce$_3$Pt$_4$In$_{13}$[32] crystals were grown instead, showing that Zn is necessary to stabilize CePtIn$_4$ crystals for the growth conditions presented here. In addition, while the previously reported YNiAl$_4$-type compounds CeNiIn$_4$[1] and CePdIn$_4$[29] can both be prepared by arc-melting, attempts to synthesize CePtIn$_4$ using this method were unsuccessful. Our description of CePtIn$_4$-Zn completes the isostructural, nominally isoelectronic 3$d$, 4$d$, 5$d$ series Ce*T*In$_4$, *T*= Ni, Pd, Pt.

A single crystal of CePtIn$_4$-Zn was mounted on a quartz sample holder with each direction of the crystal aligned parallel to the applied magnetic field to determine the direction-dependent magnetic susceptibility, as shown in **Figure 3a**. The maximum in the susceptibility ($\chi$) differs only slightly for each direction ($\chi_{max}$ = ~0.1 emu/Oe-mol$_{Ce}$ for H // $b$, $\chi_{max}$ = ~0.095 emu/Oe-mol$_{Ce}$ for H // $a$, $\chi_{max}$ = ~0.08 emu/Oe-mol$_{Ce}$ for H // $c$). The observed magnetic ordering temperature with the applied field along both the *a* and *b* direction is 1.9 K (**Figure 3a** upper left inset).

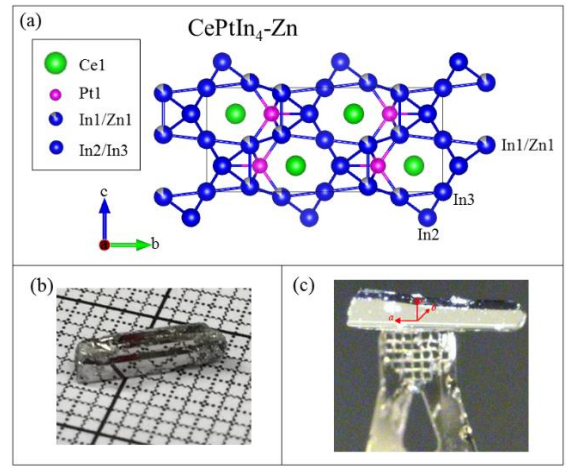

**Figure 2**. (a) Crystal structure of CePtIn$_4$-Zn, i.e. CePtIn$_4$ stabilized by 10 % Zn substituted for In. Cerium is shown in green, platinum in pink, indium in blue, and zinc in grey. (b) A typical brick-shaped crystal (1.5 mm x 2 mm x 8 mm) from the flux growth of CePtIn$_4$-Zn, chemical formula CePt(In$_{0.9}$Zn$_{0.1}$)$_4$, shown on 1 mm grid paper. (c) The crystallographic directions of the flux grown crystals determined by single crystal X-ray diffraction using the face-index method.

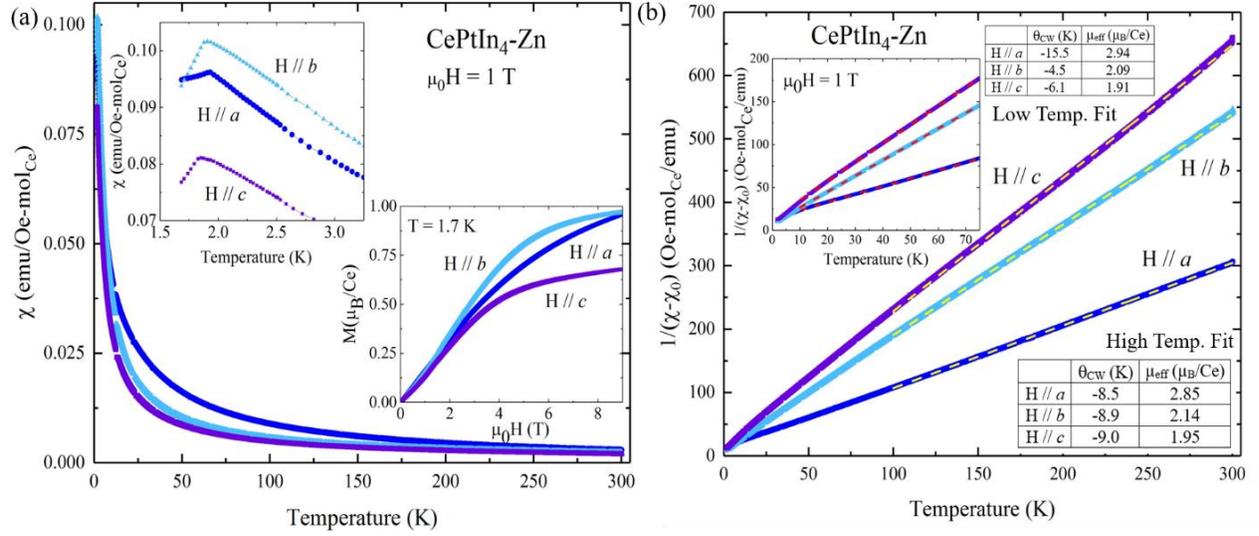

**Figure 3**. Anisotropic magnetic susceptibility for Zn-stabilized CePtIn$_4$ (CePt(In$_{0.9}$Zn$_{0.1}$)$_4$). (a) Direction-dependent magnetic susceptibility ($\chi$) in a 1 T applied magnetic field for a single crystal of CePtIn$_4$-Zn from 1.68 K – 300 K for an applied magnetic field parallel to directions *a* (blue circles), *b* (light blue triangles), and *c* (purple squares). The upper left inset plots $\chi$ vs. T at low temperatures to show the magnetic transition at ~1.9 K for each direction. The lower right inset shows the field-dependent magnetization at 1.7 K with a field sweep from 0 to 9 T for each crystallographic direction parallel to the applied field. (b) Inverse susceptibility plotted as a function of temperature where the high temperature data (100 K – 300 K; yellow dashed line) were fitted to the Curie-Weiss law in all directions (main panel). The upper left inset shows that the low temperature data (10 K – 75 K; red dashed line) were also fitted to the Curie-Weiss law in all directions and the resulting effective moments and $\theta_{CW}$ are shown. The susceptibility plot contains a $\chi_0 = 3.4 \times 10^{-4}$ emu/Oe-mol$_{Ce}$ subtracted from the raw data for H // *b*, a $\chi_0 = 5.1 \times 10^{-4}$ emu/Oe-mol$_{Ce}$ contribution for H // *c*, and a $\chi_0 = -3.6 \times 10^{-4}$ emu/Oe-mol$_{Ce}$ contribution for H // *a*.

The ordering temperature appears to be slightly lower (1.8 K) for the field aligned along the *c* direction. The lower right inset of **Figure 3a** shows that the field-dependent magnetic susceptibilities for fields applied in each of the principal directions at 1.7 K are similar to one another and do not display hysteresis. There is also a slope change at ~ 4 T for all directions but is the most well developed for the field applied along *c*, suggesting a possible metamagnetic transition. However, since the 1.7 K temperature of the measurement is very close to the AFM ordering temperature at 1.9 K, the feature is blurred. **Figure 3b** shows the inverse susceptibilities, which follow the Curie-Weiss law $\chi - \chi_0 = \frac{C}{T - \theta_{CW}}$, for each direction fitted in both the low temperature (inset) and high temperature regimes (main panel). The calculated effective moments for both fits were slightly lower than the ideal value for Ce$^{3+}$ along the *b*-direction (low temp. fit, 2.09 $\mu_B$/Ce; high temp. fit, 2.14 $\mu_B$/Ce) and *c* direction (low temp. fit, 1.91 $\mu_B$/Ce; high temp. fit, 1.95 $\mu_B$/Ce) and slightly higher along the *a* direction (low temp. fit, 2.94 $\mu_B$/Ce; high temp. fit, 2.85 $\mu_B$/Ce). However, the average effective moment taken across all three directions was 2.30 μ$_B$/Ce for the high temperature fit and 2.29 μ$_B$/Ce for the low temperature fit.

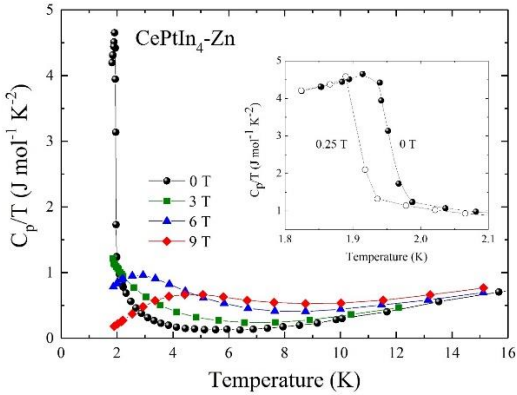

**Figure 4.** Heat capacity data of CePtIn$_4$-Zn (CePt(In$_{0.9}$Zn$_{0.1}$)$_4$) in applied fields of 0 T (black circles), 3 T (green squares), 6 T (blue triangles), 9 T (red diamonds). At higher fields (3-9T) the AFM transition peak is suppressed and a broad hump of Schottky anomaly appears shifted to higher temperatures, due to Zeeman splitting. The inset shows C$_p$/T vs T for CePtIn$_4$-Zn with applied fields of 0 T (black circles) and 0.25 T (open black circles). The peak maximum in the heat capacity shifts to a lower temperature with the application of a 0.25 T external magnetic field, consistent with AFM behavior.

The low temperature specific heat was measured both with and without applied magnetic fields, as shown in **Figure 4**. The temperature of the peak in the specific heat with no applied field is consistent with the magnetic ordering temperature observed in the χ vs. T plots. When a 0.25 T magnetic field was applied (**Figure 4** inset), the peak shifted to lower temperature, consistent with expectations for a largely AFM character of the magnetic ordering. As the field is increased to 3T, 6T, and 9T, the ordering temperature is further suppressed. Above 3T the AFM transition is suppressed and a broad hump emerges at ~4 K, which is then shifted towards higher temperatures with increasing magnetic field. This feature can be attributed to a Schottky anomaly resulting from splitting of the crystal-field split 4f energy levels and its magnetic field dependence is ascribed to the Zeeman effect.

The temperature-dependent resistivity was measured along the *a*-axis of a single crystal of CePtIn$_4$-Zn, as shown in **Figure 5**. The material has a resistivity of 58 μΩ-cm at 270 K and a residual resistivity value of 2.6 μΩ-cm at 3.5 K. The residual resistivity ratio (RRR = ρ$_{270K}$/ρ$_{3.5K}$) is ~22, suggesting that the single crystals are reasonably good metals in spite of the Zn-In disorder on the In1 site. There is a small sharp drop in the resistivity at ~3.4 K, which can be attributed to trace amounts of remnant indium on the surface of the crystal from the flux. The resistivity was also measured down to lower temperatures (**Figure 5** inset) and the compound is shown not to be superconducting down to 0.14 K. There is no clear feature of a change in resistivity at 1.9 K that can be associated with the ordering transition.

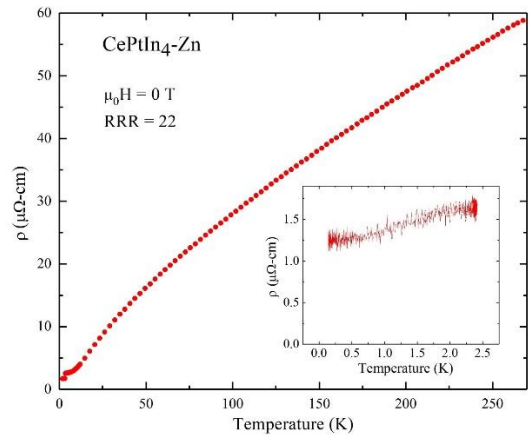

**Figure 5.** Temperature-dependent resistivity for a single crystal of CePtIn$_4$-Zn (CePt(In$_{0.9}$Zn$_{0.1}$)$_4$) taken from 1.8 K – 270 K (red circles). The inset shows temperature-dependent resistivity from 0.14 K – 2.4 K: CePtIn$_4$-Zn displays normal metallic behavior down to 0.14 K.

## 4. Conclusions

Single crystals of CePtIn$_4$-Zn were grown using a flux method, while attempts to make CePtIn$_4$ without Zn present, at ambient pressure, either by flux growth or arc-melting, were unsuccessful. CePtIn$_4$-Zn is orthorhombic and adopts the YNiAl$_4$ crystal structure. EDS data and results from the Rietveld refinement are consistent and yield the formula CePt(In$_{0.9}$Zn$_{0.1}$)$_4$ for the grown crystals, indicating that 10 % Zn stabilizes the new phase and allows for the growth of crystals of a size suitable for property measurements. Structure refinement shows that all the Zn is on the In1 position. The direction-dependent magnetic susceptibility is anisotropic, although not dramatically so, and a magnetic ordering transition is observed at 1.8-1.9 K; the structural disorder due to the presence of the 10 % Zn does not smear out the transition. Heat capacity measurements indicate that the magnetic transition at 1.9 K is sensitive to the applied magnetic field. Field-dependent magnetization measurements indicate the presence of some kind of magnetic transition at 4 T and 1.7 K. No superconductivity was observed in CePtIn$_4$-Zn down to 0.14 K. Further work to fully characterize the direction and field dependent magnetic phase diagram for CePtIn$_4$-Zn and the other members of the Ce$T$In$_4$ series may be of future interest.

**Acknowledgments** The authors thank F.A. Cevallos and J. Frick for their contributions to this work. The materials synthesis was supported by the Department of Energy, Division of Basic Energy Sciences, Grant DE-FG02-98ER45706, and the property characterization was supported by the Gordon and Betty Moore Foundation EPiQS program, Grant GBMF-4412. The work at LSU was supported by start-up funding through the LSU-College of Science. The work in Poland was supported by the National Science Centre (Poland), Grant No. UMO-2016/22/M/ST5/00435.

**Author Correspondence**
^E.M.C (carnicom@princeton.edu)
^R.J.C. (rcava@exchange.princeton.edu)

## 5. References
1. W. Schnelle, R. K. Kremer, R.-D. Hoffmann, U. C. Rodewald, and R. Pöttgen, Zeitschrift Für Naturforsch. B **69b**, 1003 (2014).
2. M. Lukachuk, Y. V. Galadzhun, R. I. Zaremba, M. V. Dzevenko, Y. M. Kalychak, V. I. Zaremba, U. C. Rodewald, and R. Pöttgen, J. Solid State Chem. **178**, 2724 (2005).
3. M. Leblanc, G. Ferey, and R. De Pape, Cryst. Growth Des. **13**, 4285 (2013).
4. P. Kushwaha, A. Thamizhavel, A. K. Nigam, and S. Ramakrishnan, Cryst. Growth Des. **14**, 2747 (2014).
5. M. D. Koterlin, B. S. Morokhivski, I. D. Shcherba, and Y. M. Kalychak, Phys. Solid State **41**, 1759 (1999).
6. V. I. Zaremba, U. C. Rodewald, R.-D. Hoffmann, Y. M. Kalychak, and R. Pöttgen, Z. Anorg. Allg. Chem. **629**, 1157 (2003).
7. S. Sarkar, M. J. Gutmann, and S. C. Peter, Dalt. Trans. **43**, 15879 (2014).
8. R. Pöttgen, J. Mater. Chem. **5**, 769 (1995).
9. I. Muts, V. I. Zaremba, V. V Baran, and R. Pöttgen, Z. Naturforsch. **62b**, 1407 (2007).
10. Y. M. Kalychak, V. I. Zaremba, Y. V Galadzhun, K. Y. Miliyanchuk, R. Hoffmann, and R. Pöttgen, Chem. Eur. J. **7**, 5343 (2001).
11. R. D. Hoffmann, R. Pöttgen, V. I. Zaremba, and Y. M. Kalychak, Z. Naturforsch. **55**, 834 (2000).
12. Y. V Galadzhun and R. Pöttgen, Z. Anorg. Allg. Chem. **625**, 481 (1999).
13. Y. Jia, C. Belin, M. Tillard, L. Lacroix-Orio, D. Zitoun, and G. Feng, Inorg. Chem. **46**, 4177 (2007).
14. R. Wawryk, J. Stępień -Damm, Z.

Henkie, T. Cichorek, and F. Steglich, J. Phys. Condens. Matter **16**, 5427 (2004).
15. M. L. Fornasini, R. Raggio, and G. Borzone, Z. Krist. **219**, 75 (2004).
16. W. D. Hutchison, D. J. Goossens, K. Nishimura, K. Mori, Y. Isikawa, and A. J. Studer, J. Magn. Magn. Mater. **301**, 352 (2006).
17. T. Mizushima, Y. Isikawa, A. Maeda, K. Oyabe, K. Mori, K. Sato, and K. Kamigaki, J. Phys. Soc. Jpn. **60**, 753 (1991).
18. R. Pöttgen, R. Müllmann, B. D. Mosel, and H. Eckert, J. Mater. Chem. **6**, 801 (1996).
19. K. M. Poduska, F. J. DiSalvo, and V. Petříček, J. Alloy. Compd. **308**, 64 (2000).
20. R. Settai, T. Takeuchi, and Y. Onuki, J. Phys. Soc. Jpn. **76**, 1 (2007).
21. C. Petrovic, P. G. Pagliuso, M. F. Hundley, R. Movshovich, J. L. Sarrao, J. D. Thompson, Z. Fisk, and P. Monthoux, J. Phys. Condens. Matter **13**, L337 (2001).
22. R. Movshovich, M. Jaime, J. D. Thompson, C. Petrovic, Z. Fisk, P. G. Pagliuso, and J. L. Sarrao, Phys. Rev. Lett. **86**, 5152 (2001).
23. D. T. Adroja, S. K. Malik, B. D. Padalia, and R. Vijayaraghavan, Phys. Rev. B **39**, 4831 (1989).
24. R. Hauser, H. Michor, E. Bauer, G. Hilscher, and D. Kaczorowski, Phys. B **230**–**232**, 211 (1997).
25. Y. Onuki, Y. Inada, H. Ohkuni, R. Settai, N. Kimura, H. Aoki, Y. Haga, and E. Yamamoto, Phys. B **280**, 276 (2000).
26. H. Shishido, N. Nakamura, T. Ueda, R. Asai, A. Galatanu, E. Yamamoto, Y. Haga, T. Takeuchi, Y. Narumi, T. C Kobayashi, K. Kindo, K. Sugiyama, T. Namiki, Y. Aoki, H. Sato, and Y. Ōnuki, J. Phys. Soc. Jpn. **73**, 664 (2004).
27. D. V. Shtepa, S. N. Nesterenko, A. I. Tursina, E. V. Murashova, and Y. D. Seropegin, Moscow Univ. Chem. Bull. **63**, 162 (2008).
28. S. N. Nesterenko, A. I. Tursina, D. V. Shtepa, A. V. Gribanov, H. Noel, and Y. D. Seropegin, J. Alloy. Compd. **442**, 93 (2007).
29. K. Momma and F. Izumi, J. Appl. Cryst. **44**, 1272 (2011).
30. R. Hoffmann and R. Pöttgen, Chem. Eur. J. **6**, 600 (2000).
31. M. F. Hundley, J. L. Sarrao, J. D. Thompson, R. Movshovich, M. Jaime, C. Petrovic, and Z. Fisk, Phys. Rev. B **65**, 24401 (2001).